\documentclass[superscriptaddress,amsmath,amssymb,aps,prl,reprint]{revtex4-1}

\usepackage{graphicx}
\usepackage{dcolumn}
\usepackage{bm}
\usepackage{hyperref}
\usepackage{notes2bib}
\usepackage{siunitx}
\usepackage{amsmath}
\usepackage[usenames,dvipsnames]{xcolor}
\usepackage[utf8]{inputenc}

\setcitestyle{super}

\begin{document}

\preprint{APS/123-QED}

\title{Observation of universal thermopolarization effect in insulators}

\author{Shuichi Iwakiri}
\email{iwakiri.shuichi@nims.go.jp}
\affiliation{
 National Institute for Materials Science, Tsukuba, Japan
}

\author{Yasumitsu Miyata}
\affiliation{
 National Institute for Materials Science, Tsukuba, Japan
}

\author{Takao Mori}
\affiliation{
 National Institute for Materials Science, Tsukuba, Japan
}

\date{\today}

\begin{abstract}
Heat-to-charge conversion has traditionally been realized via the Seebeck effect in conductors and pyroelectricity in polar insulators. 
Here, we demonstrate that temperature gradients generate electrical polarization, namely thermopolarization, in a wide range of insulators through a thermomechanical pathway.
We identify a mechanism where thermal expansion under a temperature gradient produces strain gradients that induce polarization via the flexoelectric effect.
Using a device with an on-chip heater, we detect the heat-induced polarization in crystalline, polymeric, and amorphous systems, including MgO, Al$_2$O$_3$, MnO, mica, PET, PEN, polyimide, and soda-lime glass. 
The magnitude of the response exhibits a robust scaling with the coefficient of thermal expansion, which is reproduced by finite-element simulations.
Furthermore, we identify two routes to enhance the response: reducing the sample thickness and exploiting structural instabilities such as glass and antiferromagnetic phase transitions, where more than an order-of-magnitude enhancement is observed.
These results establish a symmetry-independent route for heat-to-charge conversion in insulators and provide a device-compatible platform for electrically probing lattice responses, with potential for enhancement in nanoscale systems such as two-dimensional materials.
\end{abstract}

\maketitle
\begin{figure}[h]
\includegraphics[width=\columnwidth]{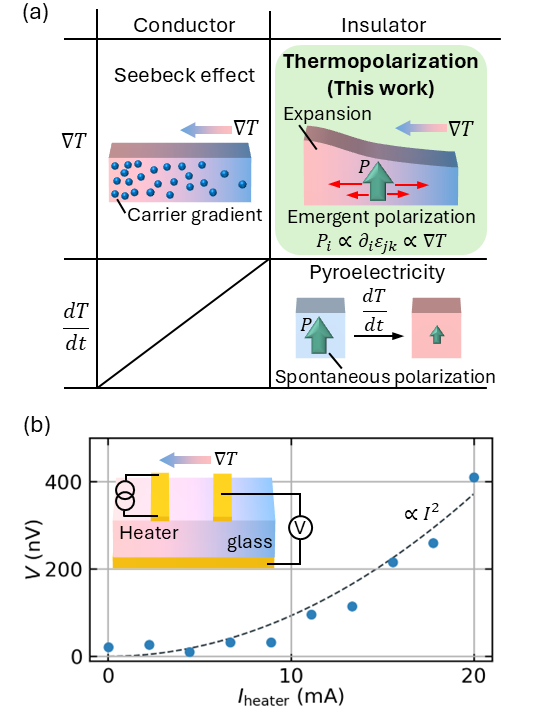} \caption{(a) Comparison of various mechanism of heat-to-charge conversion. Left: the Seebeck effect. Top right: thermopolarization (this work). Bottom Right: pyroelectricity. (b) Demonstration of the voltage generation across a glass sample with applied temperature gradient.} 
\label{concept} 
\end{figure}

\textit{Introduction--}
Thermoelectricity, or heat-to-charge conversion, is a fundamental phenomenon in condensed matter physics with broad technological importance.
In conductors, temperature gradients universally generate electrical currents through the Seebeck effect (Fig.~\ref{concept}(a) left), which serves as a powerful probe of entropy and electronic correlations in quantum materials
\cite{Mahapatra2020_crossplane_thermoelectric_TBG,Paul2022_giant_thermopower_TBG,Mravlje2016PRL,Behnia2004_thermoelectric_correlated_electrons}.

In insulators, where charge transport is absent, the established route for heat-to-charge conversion is pyroelectricity (Fig.~\ref{concept}(a) bottom right). Here, temporal temperature variations induce electrical currents via changes in polarization, which can either be spontaneous polarization or induced one by external symmetry breaking such as interfaces or strain
\cite{Liu2018_mechanisms_pyroelectricity,Masuki2023_quasiharmonic_pyroelectricity_PRB,Liu2016_firstprinciples_pyroelectricity,Yang2020_interface_polar_symmetry,Meirzadeh2019SurfacePyroSTO,Gao2026_flexo_pyroelectric_effect}. 
Beyond pyroelectricity, whether a temperature gradient can directly generate electrical polarization has been an active topic of research, often referred to as thermopolarization.
Various forms of thermopolarization have been proposed, including electronic polarization responses in solids \cite{Onishi2025,Nasu2022_thermopolarization_toroidicity} and ionic polarization in liquids \cite{Romer2012,Bresme2008,Wirnsberger2018}. 

A particularly fundamental mechanism for insulating solids was proposed by Tagantsev \cite{Tagantsev1987}, based on the flexoelectric effect, namely the generation of electrical polarization $P_i$ by strain gradients $\partial \varepsilon_{jk} / \partial x_l$, expressed as $P_i = \mu_{ijkl} \partial \varepsilon_{jk} / \partial x_l$ ($i,j,k,l = x,y,z$). 
The flexoelectric tensor $\mu_{ijkl}$ defines the response, and this effect is symmetry-allowed in materials of any symmetry \cite{tagantsev1985,Zubko2013_flexo_review,MorenoGarcia2024_flexo_HfO2,Wen2025_flexoelectric_ice,Wen2025_streaming_flexoelectricity}.
Then, when a temperature gradient is applied to a material, the nonuniform thermal expansion generates a strain gradient in it, creating a polarization via the flexoelectric effect (Fig. \ref{concept}(a) top right). 
Although the original derivation was formulated for ionic crystals, both thermal expansion and flexoelectricity are ubiquitous in solids, suggesting a universal thermopolarization in insulating materials. 
However, this route of heat-to-charge conversion has generally been considered negligible \cite{lubomirsky2012practical} except in certain ferroelectrics and quantum paraelectrics \cite{Marvan1969,Gurevich1981,Kholkin1982,GurevichTagantsev1982,Trepakov1989,Trepakov1993,Tagantsev1991,Yurkov2019_flexoelectric_inhomogeneous_heating}, and experimental investigation of its generality has been lacking.
Therefore, whether thermopolarization universally exists in insulating solids, how it can be detected, and whether it can be enhanced remain open questions.

In this work, we experimentally demonstrate that temperature gradients generate electrical polarization in a wide range of insulating materials through the flexoelectric effect. 
Using an on-chip heater to create temperature gradients and the resulting strain gradients, we induce polarization in centrosymmetric crystalline, polymeric, and amorphous systems. 
The magnitude of the response scales with the coefficient of thermal expansion, which is quantitatively reproduced by finite-element simulations. 
We further identify two pathways for enhancing thermopolarization: reducing the sample thickness and exploiting structural instabilities such as glass and antiferromagnetic phase transitions.
These results establish an insulating analogue of the Seebeck effect and provide a route to probe materials previously regarded as electrically inactive.

\begin{figure*}
\includegraphics[width=\textwidth]{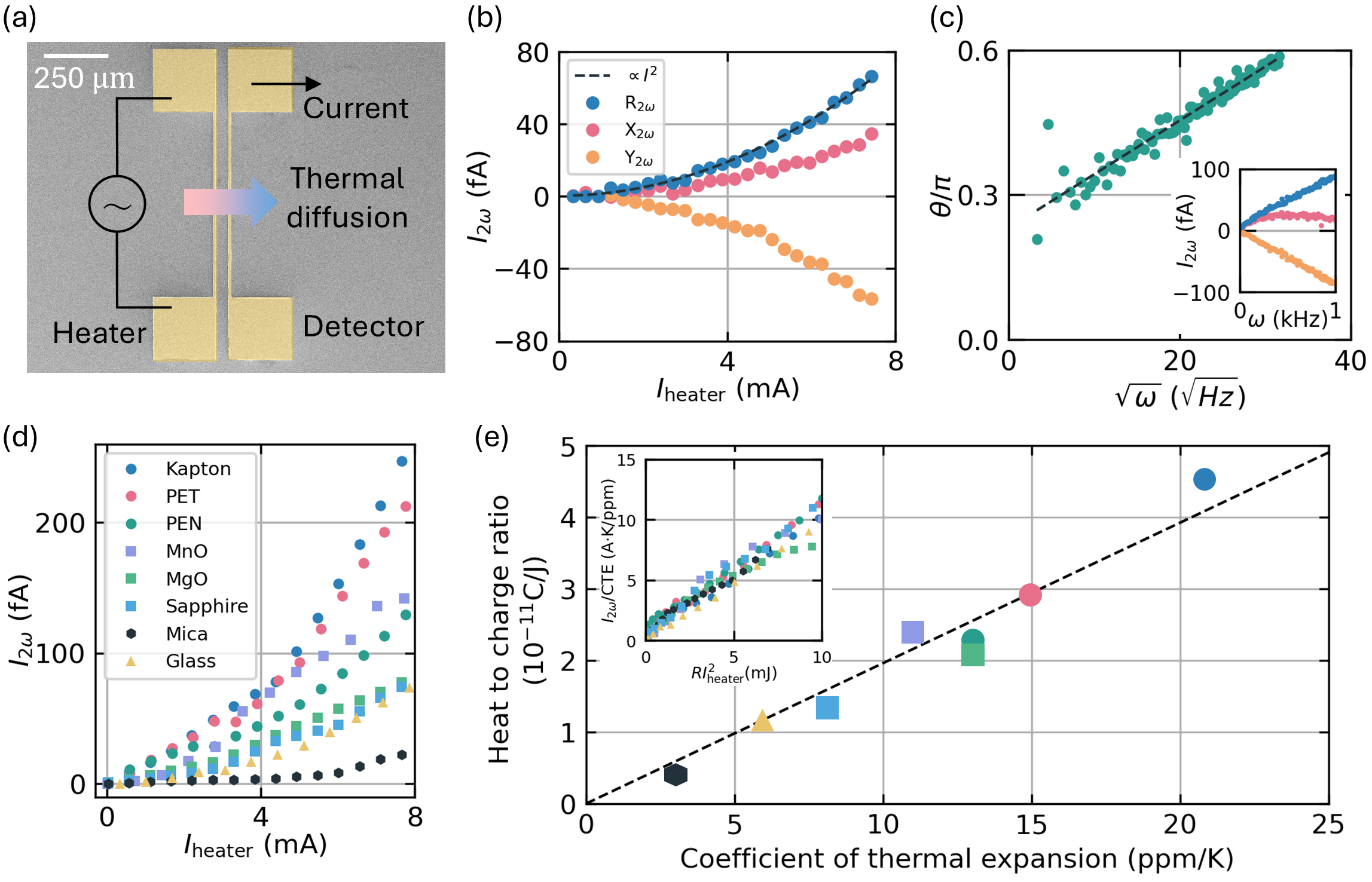}
\caption{(a) Device structure. The on-chip heater is driven by an AC current, creating a diffusive temperature wave. The generated current is detected by the detector that has the same shape as the heater. (b) Heater current dependence of the second harmonic component of the detected current. The magnitude $R_{2\omega}$, in-phase component $X_{2\omega}$, and out-of-phase component $Y_{2\omega}$ are shown. (c) Frequency dependence of the phase of the detected current. The inset shows the frequency dependence of the amplitude of the signal.(d) Heater–current dependence of $I_\mathrm{2\omega}$ for van der Waals material (mica), ionic oxide (MgO, sapphire, and MnO), polymer (PET, PEN, polyimide), and amorphous (glass).
(e) Scaling of normalized current versus coefficient of thermal expansion (CTE). Black broken line is the guide to the eye.}
\label{scaling}
\end{figure*}

\textit{Results--}
Figure~\ref{concept}(b) shows a demonstration of thermopolarization in a soda-lime glass substrate.
An on-chip heater and metallic electrodes are fabricated on the top and bottom surfaces of a 1-mm-thick substrate. 
The heater and the top detector electrode have the same geometry as those in Fig.~\ref{scaling}(a), which will be discussed in detail later.
The sample is mounted on a room-temperature stage, which fixes the bottom boundary condition.
When a current is applied to the heater, a temperature gradient is generated across the substrate. 
As a result, the temperature gradient develops both in-plane and out-of-plane, giving rise to corresponding strain gradients.

In our setup, the detected response is dominated by the out-of-plane polarization generated via the flexoelectric effect, which can be expressed as
\begin{equation}
P_z \sim \mu \frac{\partial}{\partial z}\left(\varepsilon_{xx} + \varepsilon_{yy}\right) \sim \mu \alpha \nabla T,
\label{firsteq}
\end{equation}
where $P_z$ is the out-of-plane polarization, $\varepsilon_{xx}$ and $\varepsilon_{yy}$ are the in-plane strains, $\alpha$ is the coefficient of thermal expansion (CTE), and $\nabla T$ denotes the temperature gradient. 
We also approximate the flexoelectric tensor as $\mu_{ijkl}\simeq\mu_{zjkz} \simeq \mu$, under the assumption that, for the present geometry, the measured response is predominantly governed by the out-of-plane polarization. This approximation is justified by considering that the strain gradients vary primarily along the thickness direction, while the leading strain components are in-plane.

We apply an AC current $I = I_{\mathrm{heater}} \sin \omega t$ with $\omega/2\pi = 1~\mathrm{kHz}$, which produces Joule heating and a temperature gradient
$\nabla T(t) \propto I_{\mathrm{heater}}^{2} \sin^{2}(\omega t) \propto 1 - \cos(2\omega t)$.
Then, the thermopolarization is detected as a second harmonic voltage across the sample.
We indeed observe such voltage generation as shown in Fig. \ref{concept}(b), demonstrating the presence of thermopolarization.
While the voltage generation marks a signature of thermopolarization, for a more quantitative and systematic analysis of thermopolarization, it is preferable to measure the current induced in the detector by the temporal variation of the polarization, since the voltage across the substrate is strongly influenced by the specific capacitance characteristics of the device.

Figure~\ref{scaling}(a) shows the device consisting of an on-chip heater and a metallic detector wire on a soda-lime glass. 
The heater is again driven by an alternating current
$I = I_{\mathrm{heater}} \sin \omega t$.
Under such modulation, the change in the induced polarization rather than the polarization itself generates a screening current in the detector electrode,
\begin{equation}
\frac{dP}{dt}
= \frac{dP}{dT} \frac{dT}{dt}
\propto 
\frac{d}{dT}\left(\mu \frac{\partial \varepsilon_{xx,yy}}{\partial z}\right)\frac{dT}{dt}.
\end{equation}
Since the temperature follows
$T(t) \propto I_{\mathrm{heater}}^{2} \sin^{2}(\omega t)$,
its time derivative becomes
$\frac{dT}{dt} \propto I_{\mathrm{heater}}^{2} \sin(2\omega t)$.

We observe such a current generation by measuring the second-harmonic current response $I_\mathrm{2\omega}$. 
Figure \ref{scaling}(b) shows that both the in-phase ($X$) and out-of-phase ($Y$) components of $I_{2\omega}$ increase quadratically with $I_{\mathrm{heater}}$, in agreement with the above discussion. Unless otherwise noted, all data are obtained at $\omega/2\pi = 1~\mathrm{kHz}$, and the substrate is placed on a metallic stage which is kept at room temperature. 
The heater resistance $R$ is typically 150 $\Omega$.
In addition, we conduct multiple control experiments to exclude spurious effects such as capacitive coupling and thermally-stimulated current (see END MATTER).

We also find that the lock-in phase and its frequency dependence encode the thermal origin of the generated signal. In the device, the temperature  approximately follows the one-dimensional diffusion equation $\frac{\partial T}{\partial t} = D\,\frac{\partial^{2}T}{\partial x^{2}}$ where $D$ is the thermal diffusivity. The generated heat propagates to the detector in the sample plane spending a finite time, so the temperature at distance $x$ is
$T(x,t) \propto 
\cos\!\left(2\omega t - x\sqrt{\frac{\omega}{2D}}\right)$, making the phase lag 
$\theta(\omega) = -\,x\sqrt{\frac{\omega}{2D}}\propto\sqrt{\omega}$, a hallmark of diffusive transport \cite{Angstrom1863_thermal_conductivity, Morikawa1998_high_order_harmonics, Morikawa2009_polyimide_diffusivity,Ordonez-Miranda2023_Analytical3omega}. The generated current $I_{2\omega}\propto dT/dt$ inherits the same frequency dependence of the phase.
As shown in Fig. \ref{scaling}(c), we observe that the phase $\theta(\omega)$ changes proportionally to $\sqrt{\omega}$. 
From the fitting, we find that the thermal diffusion coefficient of the glass to be $D \approx (6\pm1) \times10^{-7}~\mathrm{m^{2}/s}$, in agreement with reported values for soda-lime glass \cite{Astrath2005_thermal_lens_glass}.
This agreement between experiment and the thermal diffusion model indicates that the generated current is of thermal origin.


\begin{figure}
\includegraphics[width=\columnwidth]{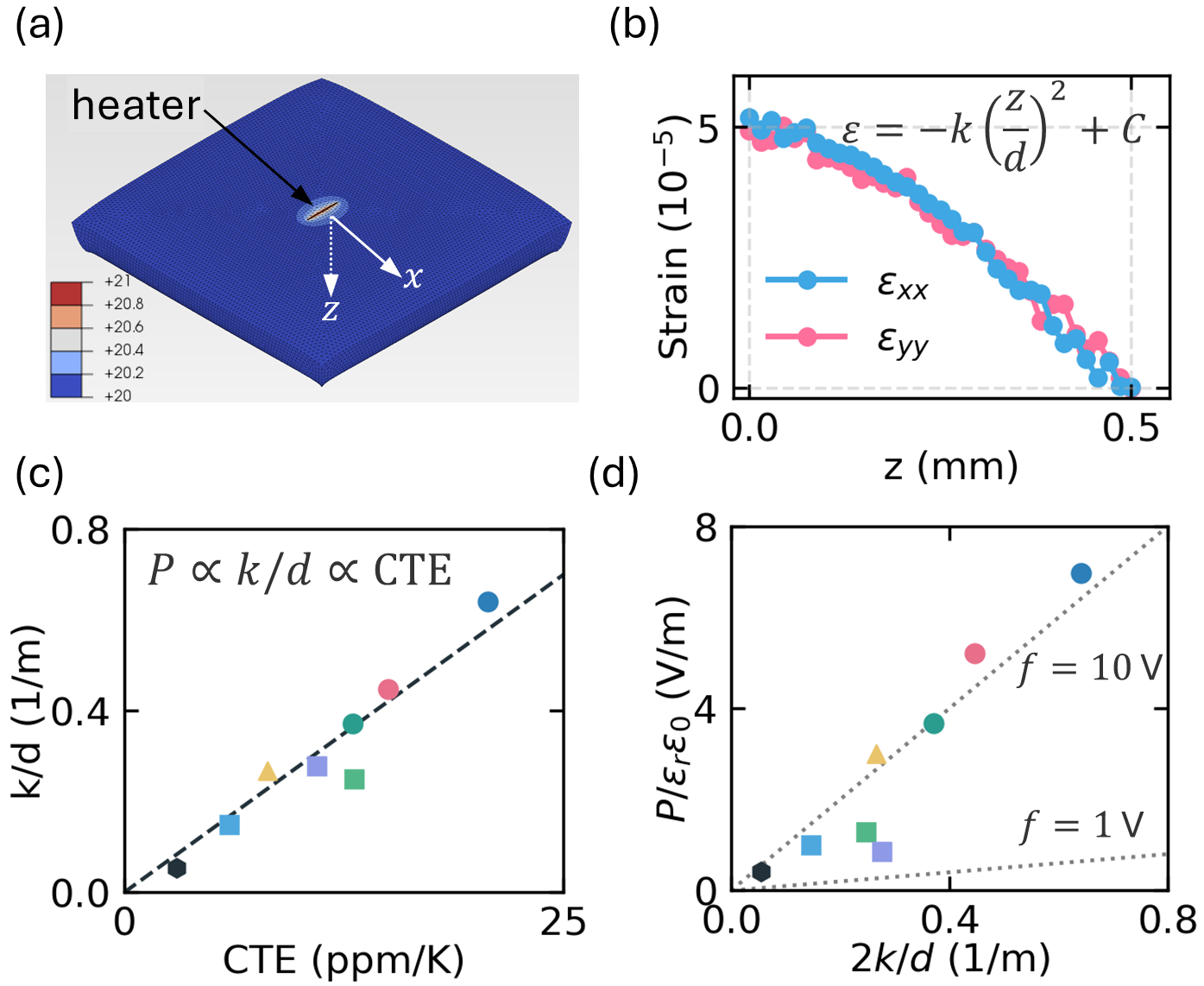}
\caption{ (a) Schematic of the FEM model. The scale bar shows the temperature in $^\circ$C. (b) Out-of-plane profile of the planar strain. (c) Coefficient $k/d$ as a function of CTE. (d) Experimentally obtained $P/\varepsilon_r \varepsilon_0$ as a function of calculated $k/d$. The broken lines illustrate the lower and upper bound of Kogan's estimate of flexocoupling coefficient $f$.
}
\label{fem}
\end{figure}

The effect is generally observed across diverse insulators.
We perform measurements on a variety of non-polar and commercially-available insulators, namely synthetic muscovite mica (0.5 mm), MgO (100, 0.5 mm), MnO (111, 0.5 mm), sapphire (0001, 0.5 mm), polyimide (Kapton, 0.25 mm), polyethylene terephthalate (PET, 0.25 mm), polyethylene naphthalate (PEN, 0.25 mm).
The values in parentheses are the crystal orientation and the thickness.
Despite their distinct bonding, structure, and thermal properties, all samples exhibit a clear second-harmonic current, as shown in Fig. \ref{scaling}(d). These results demonstrate that the thermopolarization effect is not restricted to a specific material class or crystal structure.

Moreover, we introduce a phenomenological heat-to-charge conversion ratio,
$ I_{2\omega}/(R I_\mathrm{heater}^2) $, which corresponds to the amount of polarization ($P=I_{2\omega}/\omega A$) generated per injected heating power density ($R I_\mathrm{heater}^2/\omega A$) with $A$ the electrode area.
Strikingly, all data collapse onto a single trend when plotted as a function of the coefficient of thermal expansion (CTE) $\alpha$ (ppm\,K$^{-1}$), as shown in Fig.~\ref{scaling}(e). 
We use room-temperature values of $\alpha$ for each material, as provided by the supplier.
The observed collapse indicates that the response magnitude is predominantly governed by the thermal expansion coefficient, given that the flexoelectric coefficient $\mu$ is of the same order of magnitude across the investigated materials \cite{kogan1964,tagantsev1985}.
This scaling provides direct experimental evidence for the universal nature of thermopolarization in insulators.

\textit{FEM simulations---}
Now, to validate the observed behavior, we perform finite element method (FEM) simulations.
Figure~\ref{fem}(a) shows the simulation geometry, where a heater slab is placed on top of an insulator.
The bottom plane is fixed mechanically and kept at 20 $^\circ$C.
We take the origin of the coordinate 40 $\mu$m away from the heater, which is the position of the detector. 
For simplicity, we solve the coupled thermo-mechanical equations at steady state to obtain the spatial distributions of temperature and strain in both the in-plane ($x,y$) and out-of-plane ($z$) directions.
We first perform the FEM using the parameters of MgO.
The calculated temperature decays away from the heater, as shown in Fig.~\ref{fem}(a), with the temperature modulation amplitude approximately $0.5~\mathrm{K}$.

Next, to understand the origin of the thermopolarization, we examine the strain distribution. 
Figure~\ref{fem}(b) shows the in-plane strain components $\varepsilon_{xx}$ and $\varepsilon_{yy}$ as functions of the depth $z$. According to Eq.~(\ref{firsteq}), the $z$-dependence of the strain is essential.
The simulations reveal that the strain at the detector position ($x=y=0$) varies quadratically with depth, $\varepsilon(z) = -k (z/d)^2 + C$, 
where $k$ characterizes the magnitude of the strain gradient and $C$ represents an offset. 
Here, the depth coordinate is normalized by the substrate thickness as $z/d$, such that $k$ and $C$ become dimensionless.


The calculated strain gradient can be converted to the polarization. Assuming that the sample and the detector as parallel plates, the detected polarization is written as $P=-\frac{1}{V}\int_V\mathbf{p\cdot n}dV$ \cite{Kretschmer1979_surface_effects_ferroelectrics}. Here, $\mathbf{p\cdot n}$ is the $z$ component of the flexoelectric polarization $\mathbf{p\cdot n}=p_z=\mu\frac{\partial \varepsilon_{xx,yy}}{\partial z}$. By plugging in the relation $\varepsilon=-k(z/d)^2+C$ and performing the integration across the sample volume with detector area $A$ and thickness $d$, we obtain the detected polarization as follows
\begin{equation}
P \simeq -\frac{2}{Ad}\int_0^d\mu \left(\frac{-2kz}{d}\right)\,Adz = \frac{2\mu k}{d}.
\label{polarization}
\end{equation}
The factor 2 in the right hand side of Eq. (\ref{polarization}) stems from the sum of the strain gradients in $x$ and $y$ directions.

Using the scheme described above, we perform the same simulations for all materials investigated in this work. 
The coefficient $k$, obtained from the FEM calculations, characterizes the magnitude of the strain gradient and is proportional to the generated polarization according to Eq. (\ref{polarization}). 
In Fig. \ref{fem}(c), we compare the calculated strain gradient $k/d$ with the CTE values used in the simulation. Despite the large differences in material parameters, we find that the calculated $k/d$ scales linearly with $\alpha$.
This result is consistent with the experimental scaling shown in Fig.~\ref{scaling}(e).

Furthermore, we combine the experimental results with the strain gradients derived from the simulations to compare the results with the theoretical value of the flexoelectric coefficient.
According to Kogan's estimate, the flexoelectric coefficient can be written as
$\mu = f \varepsilon_r \varepsilon_0$, where $\varepsilon_r$ is the relative permittivity, $\varepsilon_0$ is the vacuum permittivity, and $f$ is the flexoelectric coupling coefficient, typically of the order of $1$--$10$~V \cite{kogan1964}. 
This estimate provides a quantitative benchmark for testing whether the experimentally observed thermopolarization is consistent with a flexoelectric origin. 
Following this relation, Eq. (\ref{polarization}) suggests $\frac{P}{\varepsilon_r\varepsilon_0}=\frac{2fk}{d}$.
In Fig. \ref{fem}(d), we plot $P/\varepsilon_r \varepsilon_0$ as a function of $2k/d$. 
Here, the polarization $P$ is obtained from the experimentally measured current ($P \equiv I_{2\omega}/\omega A$) as shown in Fig. \ref{fem}(d). 
The result shows the ratio between $P/\varepsilon_r \varepsilon_r$ and $2k/d$ falls within Kogan's estimate.
These results confirm that the observed signals originate from flexoelectricity and quantitatively support the correlation with the thermal expansion coefficient.

\textit{Enhancement in thin films---}
According to Eq.~(\ref{polarization}), the generated polarization, and thus the detected current, is expected to scale inversely with the sample thickness $d$. Thinner samples exhibit larger strain gradients under the same thermal loading, leading to enhanced polarization.

\begin{figure}
\includegraphics[width=\columnwidth]{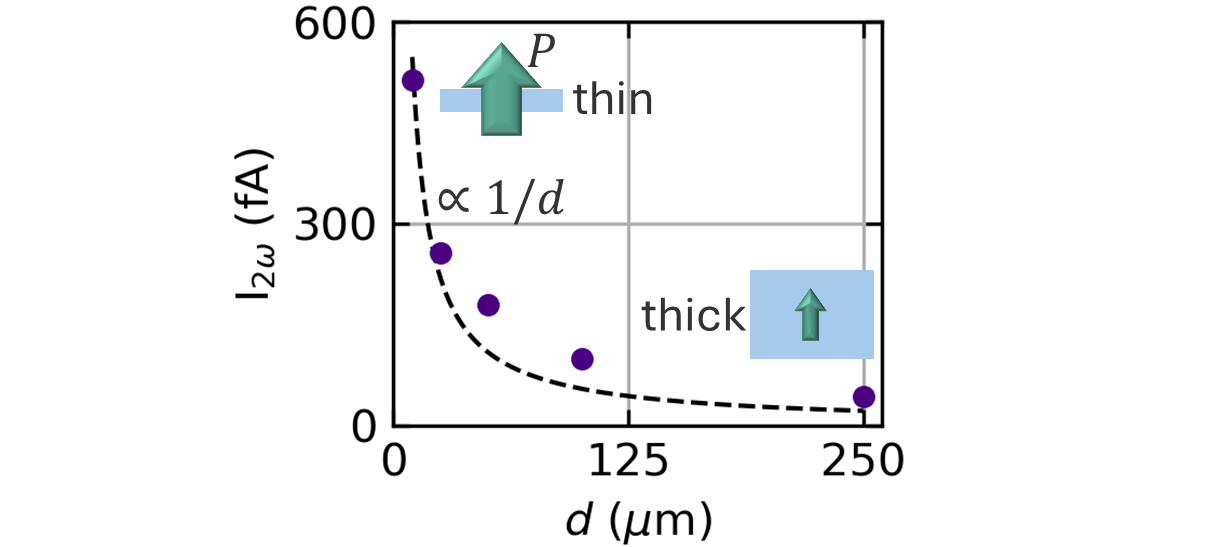}
\caption{Sample thickness $d$ dependence of the generated current in PET. The black broken curve shows $1/d$ dependence.
}
\label{ddep}
\end{figure}

We verify this by performing measurements on PET films with thicknesses $d$ of 10, 25, 50, 100, and 250~$\mu$m. The heater current is fixed at 4 mA and 1 kHz.
As shown in Fig.~(\ref{ddep}), the generated current increases markedly with decreasing thickness.
This behavior is consistent with the expected inverse thickness scaling and demonstrates that reduced dimensionality provides an effective route to enhancing thermopolarization.

\textit{Enhancement at critical points---}
\begin{figure}
\begin{center}
\includegraphics[width=\columnwidth]{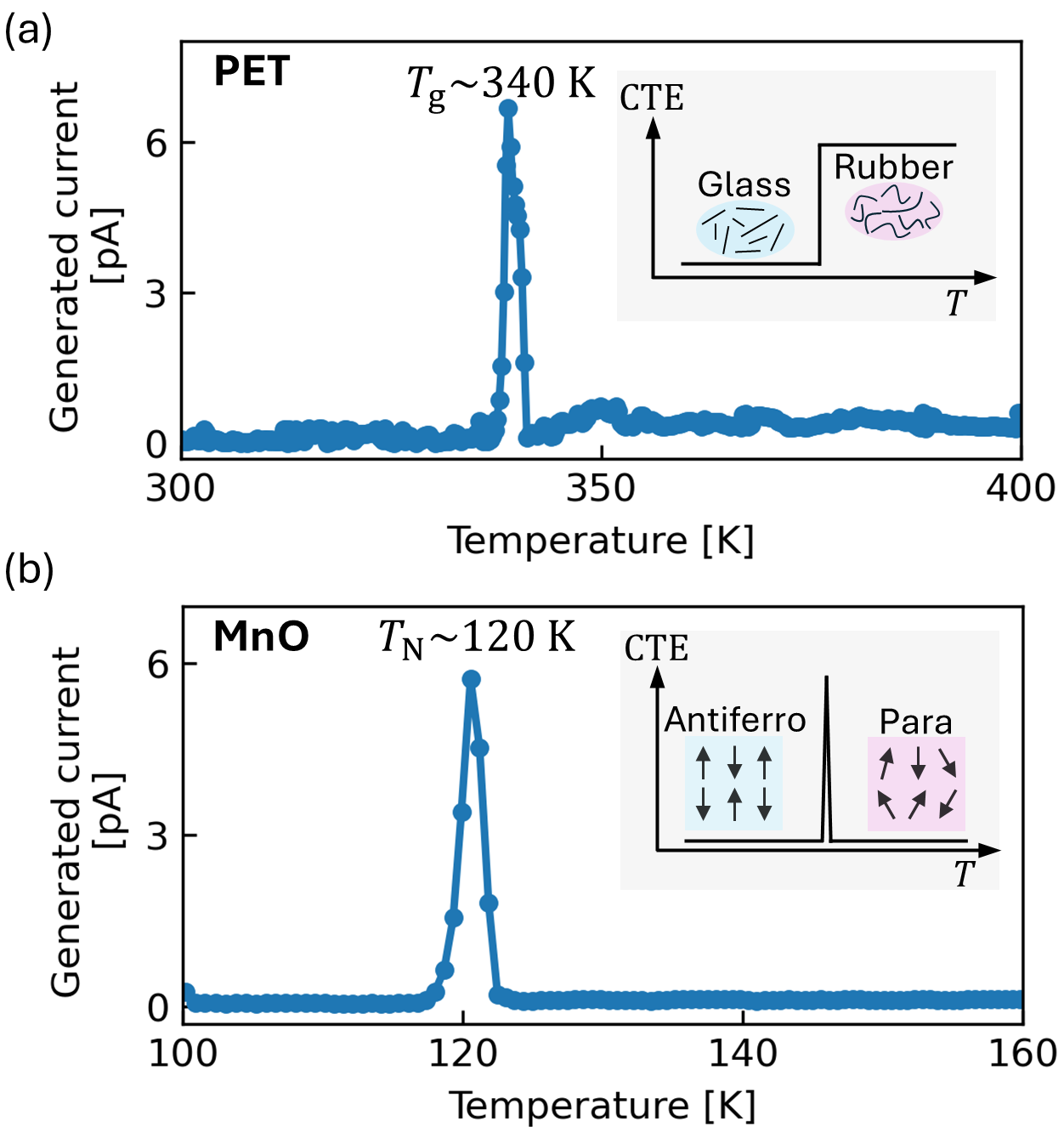}
\caption{(a) Temperature dependence of the generated current in PET around the glass transition temperature (340 K). The inset shows the schematic of the coefficient of thermal expansion (CTE) at the glass transition. (b) Temperature dependence of the generated current in MnO around the Neel temperature (120 K). The inset shows the schematic of the CTE at the Neel transition.}
\label{critical}
\end{center}
\end{figure}
Another expectation from Eq. (\ref{polarization}) is that the thermopolarization is enhanced also through the CTE, $\alpha$.
To demonstrate this, we investigate materials exhibiting critical phenomena, such as glass transitions and structural phase transitions.
Using the relation $I\propto \frac{dP}{dt}=\frac{dP}{dT}\frac{dT}{dt}\propto\frac{dP}{dT}$ and $P\sim\mu \alpha \Delta T\sim\alpha T$, we can deduce $I\propto \alpha, \frac{d\alpha}{dT}$.
Therefore, phenomena that give the coefficient of thermal expansion $\alpha$ a divergence or discontinuity would sharply enhance the thermopolarization.

Following this expectation, we first investigate the glass transition, where the CTE is known to exhibit a pronounced discontinuity \cite{Lunkenheimer2023_thermal_expansion_glass_transition} as illustrated in the inset of Fig. \ref{critical}(a).
We perform measurements on PET with heater current of 4 mA at 1 kHz. As shown in Fig.~\ref{critical}(a), a sharp enhancement of the current is observed at the glass transition temperature ($T_\textrm{g} \approx 340$~K).
%
The generated current at $T_g$ is enhanced by more than a factor of 80 compared to its room-temperature value, demonstrating a substantial amplification of thermopolarization at the glass transition.

Next, we examine a structural phase transition in an antiferromagnetic insulator such as MnO, whose transition is accompanied by a lattice distortion \cite{Bloch1973_magnetoelastic_MnO,Morosin1970_exchange_striction_MnO_MnS}.
This coupling gives rise to a divergent CTE at the magnetic transition, as schematically illustrated in the inset of Fig.~\ref{critical}(b).
Consistent with this expectation, measurements on MnO reveal a pronounced increase in the generated current at the N\'eel temperature ($T_\textrm{N} \approx 120$~K), as shown in Fig.~\ref{critical}(b).
%
The observed enhancement exceeds a factor of 70 relative to the room-temperature signal.
These results show that a structural transition, which typically is not associated with electrical transport, can nevertheless be probed electrically through our method. They further show that the thermopolarization is not a subtle effect, but can make a substantial contribution to measurements, particularly near critical points.

In summary, we demonstrate that temperature gradients universally generate electrical polarization in insulators through a thermomechanical pathway. 
The response is governed by a scaling with the thermal expansion coefficient and is strongly enhanced in reduced dimensions and near critical points associated with structural instabilities. 
This work establishes a general and engineerable route for heat-to-charge conversion in insulators previously regarded as electrically inactive. 
In particular, two-dimensional antiferromagnetic insulators, such as CrCl$_3$ \cite{Cai2019_CrCl3_antiferromagnet}, RuCl$_3$ \cite{Banerjee2016_kitaev_spin_liquid}, and FePS$_3$ \cite{Lee2016_Ising_FePS3} provide promising platforms, where reduced thickness and their structural phase transitions can lead to giant thermopolarization.

\textit{Data availability--} The data supporting the findings of this study will be available in the Figshare database.

\textit{Acknowledgments--}
The authors are grateful for the fruitful discussion with Dr. Hikaru Watanabe, Dr. Satoshi Ishii, and Dr. Jun Usami.
This work is partially supported by the Japan Society for the Promotion of Science (JSPS) KAKENHI Grant No. 25K21685, Murata Science and Education Foundation Grant No. M25AN110, Kanamori Foundation, and TIA Kakehashi Grant. T.M. thanks the support from JST Mirai JPMJMI19A1.

\bibliography{QHE_main}

\section{END MATTER}
\subsection{Control experiment: Capacitive coupling}
We rule out capacitive artifacts as the origin of the observed response.
We perform a control experiment where one of the heater pads is electrically floated. In this configuration, a voltage is applied but no current flows, eliminating Joule heating while preserving capacitive coupling. Experimentally, the second harmonic signal disappears in this configuration.

Also, the temperature dependence of capacitance would not appear in the second harmonic signal. The temperature oscillates at $2\omega$, while the voltage across the capacitor varies at $\omega$. The resulting current term scales as $V(t)\frac{dC}{dt} \propto \sin(\omega t)\sin(2\omega t)$,
which contains only $\omega$ and $3\omega$ components.

\begin{figure}[h]
\begin{center}
\includegraphics[width=\columnwidth]{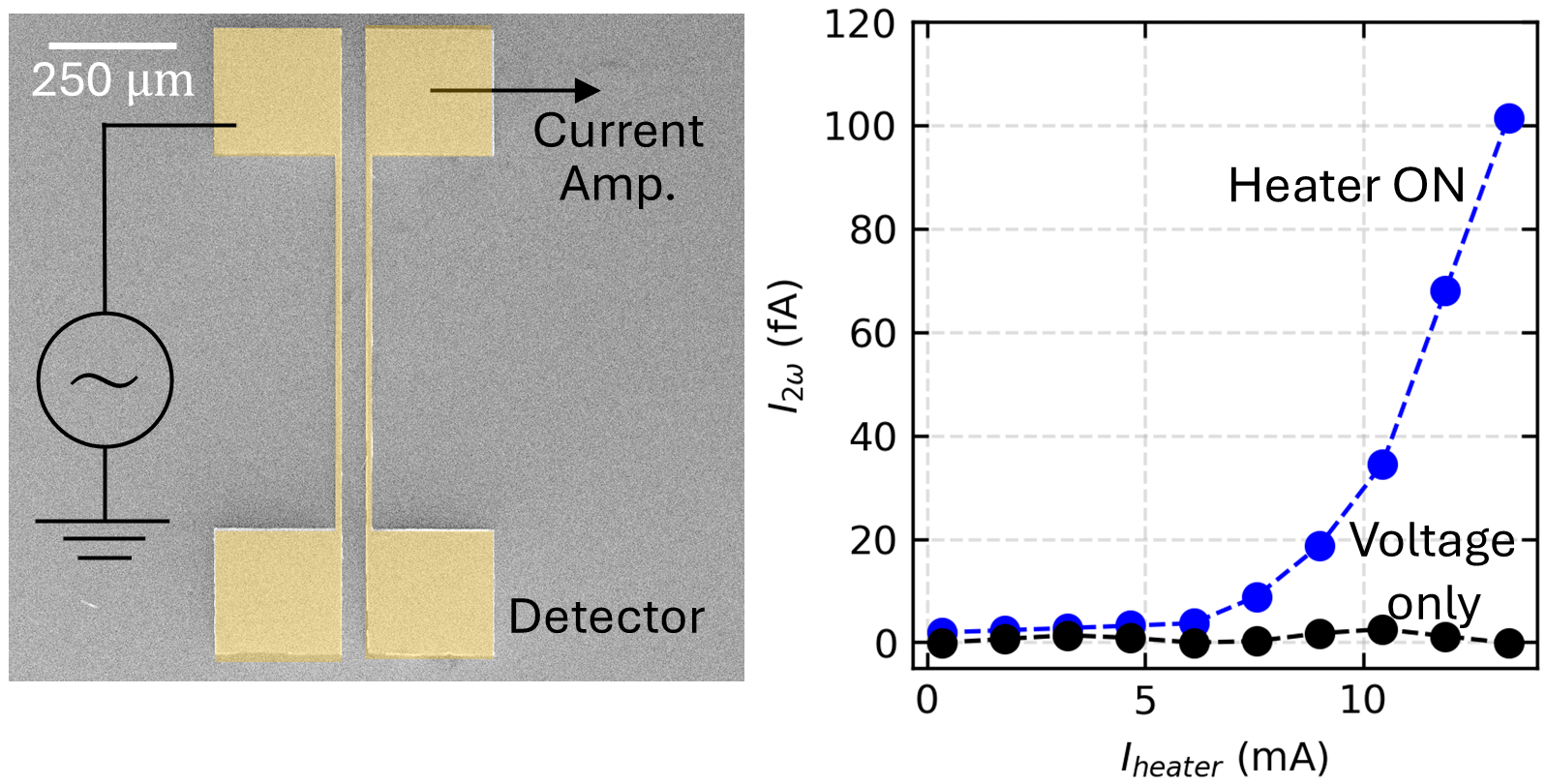}
\caption{Second-harmonic current measured under two heater configurations: (blue) normal operation with both heater terminals connected; (black) voltage-only condition in which only one heater terminal is connected.}
\label{group}
\end{center}
\end{figure}

\subsection{Time stability}
\begin{figure}[h]
\begin{center}
\includegraphics[width=\columnwidth]{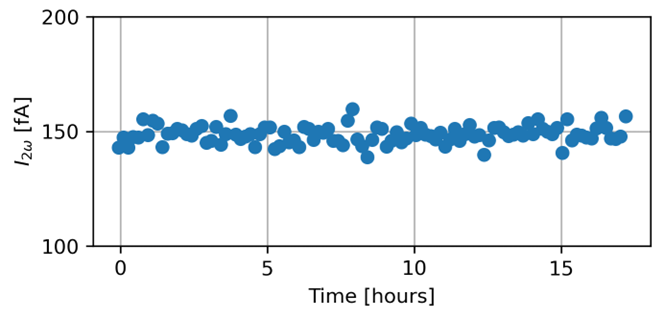}
\caption{Second-harmonic current measured for 17 hours in PET sample.}
\label{time_stability}
\end{center}
\end{figure}

The thermally stimulated current (TSC) \cite{Baldini1993_TSC_silicon_detectors,Perlman1972_TSC_TSV_dielectrics}, which arises from the release of trapped charges during heating, is also not responsible in our experiment. Such currents require prior charge trapping through bias cooling or optical excitation, which is not the case with our experiment.
We also confirm that the detector current does not decay for 17 hours (PET substrate, heater current $6~\mathrm{mA}$, excitation frequency $1~\mathrm{kHz}$) as shown in Fig.~S\ref{time_stability}. The absence of temporal decay rules out thermally stimulated currents as the origin of the observed signal.

\end{document}